Spring 3-2020

# Identifying and Mapping the Global Research Output on Coronavirus Disease: A Scientometric Study

Muneer Ahmad  
*Annamalai University*

Dr. M.Sadik Batcha  
*Annamalai University*

# Identifying and Mapping the Global Research Output on Coronavirus Disease: A Scientometric Study

**Muneer Ahmad[1] Dr. M Sadik Batcha[2]**

[1]*Research Scholar, Department of Library and Information Science, Annamalai University, Annamalai nagar, muneerbangroo@gmail.com*
[2]*Research Supervisor & Mentor, Professor and University Librarian, Annamalai University, Annamalai nagar, msbau@rediffmail.com*

**Abstract**

The paper explores and analyses the trend of world literature on "Coronavirus Disease" in terms of the output of research publications as indexed in the Science Citation Index Expanded (SCI-E) of Web of Science during the period from 2011 to 2020. The study found that 6071 research records have been published on Coronavirus Disease till March 20, 2020. The various scientometric components of the research records published in the study period were studied. The study reveals the various aspects of Coronavirus Disease literature such as year wise distribution, relative growth rate, doubling time of literature, geographical wise, organization wise, language wise, form wise , most prolific authors, and source wise. The highest number of articles was published in the year 2019, while lowest numbers of research article were reported in the year 2020. Further, the relative growth rate is gradually increases and on the other hand doubling time decreases. Most of the research publications are published in English language and most of the publications published in the form of research articles. USA is the highest contributor to the field of Coronavirus Disease literature.

***Keywords:*** COVID-19, Coronavirus, SARS-CoV-2, Wuhan Epidemic, VOSviewer, Histcite, Web of Science.

**Introduction**

Since a cluster of unidentified pneumonia patients was found in December 2019 in Wuhan, China, a new Coronavirus (CoV), which was momentarily named 2019 novel coronavirus (2019-nCoV) by the World Health Organization (WHO) on January 7, 2020, unexpectedly came into our prospect (Huang et al., 2020)**.** The virus was consequently renamed Severe Acute Respiratory Syndrome Coronavirus 2 (SARS-CoV-2), and the disease it causes was named Coronavirus Disease 2019 (COVID-19). As of March 27, 2020, there have been more than 566,269 patients confirmed positive by nucleic acid testing in China and 200 other countries, areas or territories and it has caused 25,423 deaths due to acute respiratory failure or other related complications. In addition, more than 391,904 currently infected patients were isolated and are being treated of them 371935 (95%) are in mild condition and 19,969 (5%) patients are in serious or critical condition. On January 31, WHO announced the explosion of COVID-19 in China as a Public Health Emergency of International Concern. In 2002-2003, more than 8000 patients suffered from Severe Acute Respiratory Syndrome (SARS) due to a coronavirus, with 774 virus associated deaths reported to WHO. Since September 2012, there were 2494 laboratory-confirmed cases of contagion with Middle East Respiratory Syndrome Coronavirus (MERS-CoV),with 858 virus-related deaths reported to WHO (World Health Organization, 2004, 2013). All 3 of these rising infectious diseases leading to a global spread are caused by β-coronaviruses.

Coronavirus disease (COVID-19) is an infectious disease caused by a recently discovered Coronavirus. Most people infected with the COVID-19 virus will familiarity mild to moderate respiratory illness and get well without requiring extraordinary treatment. Older people and those with fundamental health problems like cardiovascular disease, diabetes, chronic respiratory disease, and cancer are more likely to widen severe illness. At this time, there are no explicit vaccines or treatments for COVID-19. However, there are many constant clinical trials evaluating impending treatments (World Health Organization, 2020).

SARS-CoV-2 is intimately associated to two bat-derived severe acute respiratory syndrome-like coronaviruses, bat-SL-CoVZC45 and bat-SL-CoVZXC21. It is spread by human-to-human diffusion via droplets or direct contact, and infection has been projected to have incubation period of 2-14 days, however, a case with and incubation period of 27 days has been reported by Hubei Province local government on 22 February 2020. Mean incubation period observed in travellers from Wuhan 6.4 days (range from 2.1 to 11.1 days).

The COVID-19 virus affects different people in different ways. COVID-19 is a respiratory disease and most contaminated people will develop placid to moderate symptoms and pick up without requiring extraordinary treatment. People who have primary remedial circumstances and those over 60 years old have a higher risk of mounting severe disease and death. Common symptoms comprise: fever, tiredness, dry cough. Other symptoms include: shortness of breath, aches and pains, sore throat, and very few people will report diarrhea, nausea or a runny nose.

In China, prior outbreaks of emerging infections have had an inauspicious impact on the blood supply (Shan & Zhang, 2004). However, reflection must also be given to the safety of the transfusion receiver even if the emerging infection is a respiratory disease. Previous studies indicated that viral RNA could be detected from plasma or serum of patients infected with SARS-CoV (Drosten et al., 2003; Grant et al., 2003; Ng et al., 2003), MERS-CoV (Corman et al., 2015), or SARS-CoV-2 (Huang et al., 2020) during different periods after the inception of symptoms. However, the finding of viral RNA by polymerase chain reaction (PCR) is not comparable to the detection of intact infectious virus. Although WHO noted in 2003 that no cases of SARS-CoV have been reported due to transfusion of blood products, there was still a speculative risk of transmission of SARS-CoV through transfusion. With more and more asymptomatic infections being originate among COVID-19 cases, blood safety is commendable of contemplation. 3.4% mortality rate has been predictable by the WHO as of March 3, 2020. In his opening remarks at the March 3 media briefing on Covid-19, WHO Director-General Dr Tedros Adhanom Ghebreyesus stated: "Globally, about 3.4% of reported cases have died. By comparison, seasonal flu generally kills far less than 1% of those infected" (World Health Organization, 2020).

Wuhan (the city where the virus originated) is the largest city in Central China, with a population of over 11 million people. The city, on January 23, shut down transport links. Following Wuhan lock down, the city of Huanggang was also positioned in quarantine, and the city of Ezhou closed its train stations. This means than 18 million people have been placed in isolation. The World Health Organization (WHO) said cutting off a city as large as Wuhan is "unprecedented in public health history" (Reuters, 2020) and praised China for its incredible dedication to segregate the virus and diminish the spread to other countries.

The novel coronavirus' case fatality rate has been expected at around 2%, in the WHO press conference held on January 29, 2020 (WorldoMeter, 2020). However, it noted that, without knowing how many were infected, it was too early to be able to put a percentage on the mortality rate figure. A prior approximation (Wang, Horby, Hayden, & Gao, 2020) had put that number at 3%. Fatality rate can change as a virus can transform, according to epidemiologists. For comparison, the case fatality rate for SARS was 10%, and for MERS 34%.

**Review of Literature**

The review, in general, provides an overview of the theory and the research literature, with a special emphasis on the literature specific to the topic of investigation. It provides support to the proposition of one's research, with ample evidences drawn from subject experts and authorities in the concerned field. The sources consulted for the review of literature here includes Scientometric studies related materials drawn from Primary periodicals.

(Batcha & Ahmad, 2017) obtained the analysis of two journals Indian Journal of Information Sources and Services (IJSS) which is of Indian origin and Pakistan Journal of Library and Information Science (PJLIS) from Pakistan origin and studied them comparatively with scientometric indicators like year wise distribution of articles, pattern of authorship and productivity, degree of collaboration, pattern of co-authorship, average length of papers, average keywords, etc and found 138 (94.52%) of contributions from IJISS were made by Indian authors and similarly 94 (77.05) of contributions from PJLIS were done by Pakistani authors. The collaboration with foreign authors of both the countries is negligible (1.37% of articles) from India and (4.10% of articles) from Pakistan.

(Ahmad, Batcha, Wani, Khan, & Jahina, 2018) studied Webology journal one of the reputed journals from Iran through scientometric analysis. The study aims to provide a comprehensive analysis regarding the journal like year wise growth of research articles, authorship pattern, author productivity, and subjects taken by the authors over the period of 5 years from 2013 to 2017. The findings indicate that 62 papers were published in the journal during the study period. The articles having collaborative nature were high in number. Regarding the subject concentration of papers of the journal, Social Networking, Web 2.0, Library 2.0 and Scientometrics or Bibliometrics were highly noted. The results were formulated through standard formulas and statistical tools.

(Batcha, Jahina, & Ahmad, 2018) has examined the DESIDOC Journal by means of various scientometric indicators like year wise growth of research papers , authorship pattern, subjects and themes of the articles over the period of five years from 2013 to 2017. The study reveals that 227 articles were published over the five years from 2013 to 2017. The authorship pattern was highly collaborative in nature.  The maximum numbers of articles (65 %) have ranged their thought contents between 6 and 10 pages.

(Ahmad & Batcha, 2019) analyzed research productivity in Journal of Documentation (JDoc) for a period of 30 years between 1989 and 2018. Web of Science a service from Clarivate Analytics has been consulted to obtain bibliographical data and it has been analysed through Bibexcel and Histcite tools to present the datasets. Analysis part deals with local and global citation level impact, highly prolific authors and their research output, ranking of prominent institution and countries. In addition to this scientographical mapping of bibliographical data is obtainable through VOSviewer, which is open source mapping software.

(Ahmad & Batcha, 2019) studied the scholarly communication of Bharathiar University which is one of the vibrant universities in Tamil Nadu. The study find out the impact of research produced, year-wise research output, citation impact at local and global level, prominent authors and their total output, top journals of publications, top collaborating countries which collaborate with the university authors, highly industrious departments and trends in publication of the university during 2009 through 2018. During the 10 years of study under consideration it indicates that a total of 3440 research articles have been published receiving 38104 citations having h-index as 68. In addition the study used scientographical mapping of data and presented it through graphs using VOSviewer software mapping technique.

(Ahmad, Batcha, & Jahina, 2019) quantitatively measured the research productivity in the area of artificial intelligence at global level over the study period of ten years (2008-2017). The study acknowledged the trends and features of growth and collaboration pattern of artificial intelligence research output. Average growth rate of artificial intelligence per year increases at the rate of 0.862. The multi-authorship pattern in the study is found high and the average number of authors per paper is 3.31. Collaborative Index is noted to be the highest range in the year 2014 with 3.50. Mean CI during the period of study is 3.24. This is also supported by the mean degree of collaboration at the percentage of 0.83 .The mean CC observed is 0.4635. Regarding the application of Lotka's Law of authorship productivity in the artificial intelligence literature it proved to be fit for the study. The distribution frequency of the authorship follows the exact Lotka's Inverse Law with the exponent á = 2. The modified form of the inverse square law, i.e., Inverse Power Law with á and C parameters as 2.84 and 0.8083 for artificial intelligence literature is applicable and appears to provide a good fit. Relative Growth Rate [Rt(P)] of an article gradually increases from -0.0002 to 1.5405, correspondingly the value of doubling time of the articles Dt(P) decreases from 1.0998 to 0.4499 (2008-2017). At the outset the study reveals the fact that the artificial intelligence literature research study is one of the emerging and blooming fields in the domain of information sciences.

(Batcha, Dar, & Ahmad, 2019) presented a scientometric analysis of the journal titled "Cognition" for a period of 20 years from 1999 to 2018. The study was conducted with an aim to provide a summary of research activity in the journal and characterize its most aspects. The research coverage includes the year wise distribution of articles, authors, institutions, countries and citation analysis of the journal. The analysis showed that 2870 papers were published in journal of Cognition from 1999 to 2018. The study identified top 20 prolific authors, institutions and countries of the journal. Researchers from USA have made the most percentage of contributions.

**Objectives**

The present manuscript aims to study the various dimensions of coronavirus research output in terms of various scientometric indicators, based on publication and citation data, derived from Web of Science database during 2011-2020. In particular, the study analyzed overall annual and cumulative growth of global publications with relative growth rate and doubling time, its share among top 20 most productive countries, publication output distribution by document type and language used for scholarly communication, productivity and citation impact of most productive institutions and authors, and leading media of communication.

**Methodology**

For the present study, the publication data was retrieved and downloaded from Web of Science database on Coronavirus research during 2011-2020. A main search strategy for global output was formulated, where the keyword such as "Coronavirus Disease, OR Coronavirus OR COVID-19" were searched within "Topic" category and further limited the search output to period 2011-2020 within "Timespan". This search strategy generated 6071 publications on Coronavirus from Web of Science database. The year of publication, citations, source wise distribution, form wise, language used for the medium of scholarly communication, institutions and authors were analyzed and displayed in tables and scientographs by using Histcite and VOSviewer respectively. The global citation scores and local citation scores were examined to identify the pattern of research contribution on Coronavirus.

**Discussion and Result**

### Evaluate the Annual Output of Publications

The global research output in coronavirus disease research cumulated to 6071 publications in 10 years during 2011-2020 and they increased from 383 in the year 2020 to 747 publications in the year 2016. The data from Table 1 reveals that the numbers of research documents published from 2011 to 2020 shows fluctuation in publication trend. According to the publication output from the Table 1 the year wise distribution of research documents, 2016 has the highest number of research documents 747 (12.30%) with 4362 (11.62%) of total local citation score and 9729 (10.09%) of total global citation score values and being prominent among the 10 years output and it stood in first rank position. The year 2014 has 715 (11.78%) research documents and it stood in second position with 8468 (22.56%) of total local citation score and 18824 (20.02%) of total global citation score were scaled. It is followed by the year 2019 with 714 (11.76 %) of records and it stood in third rank position along with 306 (0.97%) of total local citation score and 1039 (1.08%) of total global citation score measured. The year 2015 has 692 (11.40%) research documents and it stood in fourth position with 5118 (13.64%) of total local citation score and 13056 (13.53%) of total global citation score were scaled. It has been observed that increase in publications in the research hasn't direct impact on citation score. The table presents the year wise publications and depicts the citation score. It clearly indicates on the fact that increase in publication rate is not directly linked to increase in citation Score.

Table 1: Annual Distributions of Publications and Citations

| S.No. | Year | Records | % | TLCS* | % | TGCS* | % |
|---|---|---|---|---|---|---|---|
| 1 | 2011 | 409 | 6.74 | 2996 | 7.98 | 11339 | 11.75 |
| 2 | 2012 | 461 | 7.59 | 4605 | 12.27 | 13451 | 13.94 |
| 3 | 2013 | 617 | 10.16 | 8584 | 22.87 | 19313 | 20.02 |
| 4 | 2014 | 715 | 11.78 | 8468 | 22.56 | 18824 | 19.51 |
| 5 | 2015 | 692 | 11.40 | 5118 | 13.64 | 13056 | 13.53 |
| 6 | 2016 | 747 | 12.30 | 4362 | 11.62 | 9729 | 10.09 |
| 7 | 2017 | 687 | 11.32 | 1709 | 4.55 | 5992 | 6.21 |
| 8 | 2018 | 646 | 10.64 | 1018 | 2.71 | 3153 | 3.27 |
| 9 | 2019 | 714 | 11.76 | 306 | 0.82 | 1039 | 1.08 |
| 10 | 2020 | 383 | 6.31 | 363 | 0.97 | 571 | 0.59 |
| Total | | 6071 | 100.00 | 37529 | 100 | 96467 | 100 |

*TLCS = Total Local Citation Score, *TGCS = Total Global Citation Score

**Relative Growth Rate and Doubling Time**

It is very clear that the relative growth rate of total literature outputs published has been progressively improved. The growth rate is 0.64 in 2012, which is increased up to 2.76 in 2020. The mean relative growth rate is 1.44 during the period 2011-2020. Generally, the relative growth rate of publications of all sources in this data has shown an increasing trend. The mean doubling time is 0.47 during the period 2011-2020. In general, the doubling time of scholarly publications of all sources in this research output has also shown a decreasing trend.

Table 2: Relative Growth Rate and Doubling Time

| S.No. | Year | Records | Cum. No. of Records | W1 | W2 | R(a) W2-W1 | Mean R(a) (1-2) | Doubling Time Dt (a) | Mean Dt (a)(1-2) |
|---|---|---|---|---|---|---|---|---|---|
| 1 | 2011 | 409 | 409 | 6.01 | 6.01 | 0.00 | 0.81 | | 0.60 |
| 2 | 2012 | 461 | 870 | 6.13 | 6.77 | 0.64 | | 1.09 | |
| 3 | 2013 | 617 | 1487 | 6.42 | 7.30 | 0.88 | | 0.79 | |
| 4 | 2014 | 715 | 2202 | 6.57 | 7.70 | 1.12 | | 0.62 | |
| 5 | 2015 | 692 | 2894 | 6.54 | 7.97 | 1.43 | | 0.48 | |
| 6 | 2016 | 747 | 3641 | 6.62 | 8.20 | 1.58 | 2.06 | 0.44 | 0.35 |
| 7 | 2017 | 687 | 4328 | 6.53 | 8.37 | 1.84 | | 0.38 | |
| 8 | 2018 | 646 | 4974 | 6.47 | 8.51 | 2.04 | | 0.34 | |
| 9 | 2019 | 714 | 5688 | 6.57 | 8.65 | 2.08 | | 0.33 | |
| 10 | 2020 | 383 | 6071 | 5.95 | 8.71 | 2.76 | | 0.25 | |
| Total | | 6071 | | | | | 1.44 | | 0.47 |

**Publication Profile of Top 20 Most Productive Countries**

More than 120 countries of the world participated in global research in coronavirus disease research during 2011-2020. Between 88 and 2019 publications were contributed by top 20 most productive countries in coronavirus disease research. Each of the top 20 countries had global publication share between 1.40% and 33.30% during 2011-2020. USA accounted for the highest publication share (33.30%), followed by Peoples Republic of China (24.40%), UK (7.10%), Saudi Arabia (6.80%), Germany (6.70%), South Korea (5.50%), Netherlands (5.10%), France (4.90%), Japan (4.10%), and Canada (3.80%) followed by other countries. By using Country Mapping Analysis, it has been found that the nodes are linked to each other indicating that countries are having collaboration with other associated nations. It could be identified from the analysis the following countries: USA, Peoples Republic of China, UK, Saudi Arabia, Germany, South Korea, Netherlands, France, Japan, and Canada etc were identified the most productive countries based on the number of research papers published.

Table 3: Distribution of the Publication Output of Top 20 Countries

| S.No. | Country | Records | % | TLCS | TGCS |
|---|---|---|---|---|---|
| 1 | USA | 2019 | 33.30 | 15804 | 42725 |
| 2 | Peoples R China | 1481 | 24.40 | 10127 | 23444 |
| 3 | UK | 434 | 7.10 | 4697 | 12137 |
| 4 | Saudi Arabia | 411 | 6.80 | 6968 | 12263 |
| 5 | Germany | 405 | 6.70 | 5290 | 11348 |
| 6 | South Korea | 332 | 5.50 | 1650 | 3801 |

| 7 | Netherlands | 307 | 5.10 | 5605 | 11563 |
|---|---|---|---|---|---|
| 8 | France | 296 | 4.90 | 2020 | 4936 |
| 9 | Japan | 251 | 4.10 | 1066 | 2802 |
| 10 | Canada | 232 | 3.80 | 1193 | 4057 |
| 11 | Australia | 185 | 3.00 | 1206 | 3717 |
| 12 | Italy | 184 | 3.00 | 689 | 2428 |
| 13 | Switzerland | 156 | 2.60 | 1343 | 3779 |
| 14 | Spain | 144 | 2.40 | 1212 | 3171 |
| 15 | Brazil | 141 | 2.30 | 311 | 1030 |
| 16 | Taiwan | 140 | 2.30 | 473 | 1536 |
| 17 | Singapore | 137 | 2.30 | 770 | 2257 |
| 18 | Egypt | 131 | 2.20 | 1010 | 2433 |
| 19 | India | 89 | 1.50 | 89 | 629 |
| 20 | Sweden | 88 | 1.40 | 846 | 2123 |

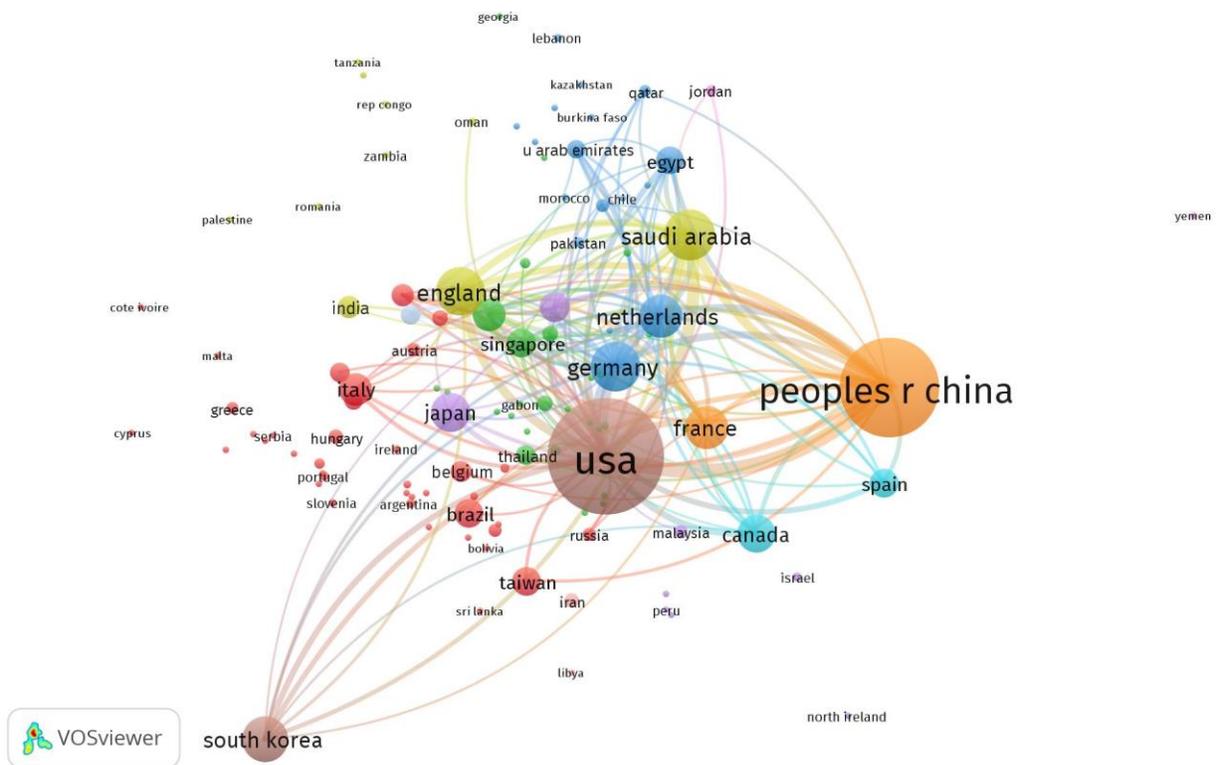

Figure 1: Countries having collaborating nodes

**Distribution of Language of Publications**

Table 4 reveals the language of publications. The research literature output in Coronavirus Disease during the period of coverage was found to be in 15 languages among which English was predominant with 98.53%. Non-English contributions belonging to other 14 languages shared 1.47% of the total output forming a meagre number. English proved to be the *lingua franca* to the scientific community engaged in coronavirus or Covid-19 research across the world. Out of the 1.47% of non-English literature, a majority was in European languages that included French, Spanish, German, Hungarian, Polish, Italian, Dutch, Portuguese and clusters around Russia. Turkish, Chinese, Czech, Greek and Slovene also figured in. There was a single article in Czech, Greek and Slovene Languages while there was not even a single one in Hindi.

Table 5: Distribution of Language of Publications

| S.No. | Language | Records | % | TLCS | TGCS |
|---|---|---|---|---|---|
| 1 | English | 5982 | 98.53 | 37507 | 96275 |
| 2 | French | 18 | 0.30 | 1 | 9 |
| 3 | Spanish | 16 | 0.26 | 8 | 35 |
| 4 | German | 14 | 0.23 | 4 | 28 |
| 5 | Hungarian | 8 | 0.13 | 1 | 4 |
| 6 | Polish | 7 | 0.12 | 1 | 3 |
| 7 | Turkish | 7 | 0.12 | 7 | 43 |
| 8 | Chinese | 4 | 0.07 | 0 | 2 |
| 9 | Italian | 4 | 0.07 | 0 | 1 |
| 10 | Dutch | 3 | 0.05 | 0 | 61 |
| 11 | Portuguese | 3 | 0.05 | 0 | 4 |
| 12 | Russian | 2 | 0.03 | 0 | 1 |
| 13 | Czech | 1 | 0.02 | 0 | 1 |
| 14 | Greek | 1 | 0.02 | 0 | 0 |
| 15 | Slovene | 1 | 0.02 | 0 | 0 |
|  |  | 6071 | 100.00 | 37529 | 96467 |

**Form Wise Analysis**

The analysis to preference sources by the productive scientists for publication output in Coronavirus Disease is an essential aspect of bibliometric and scientometric analysis. Scientists have communicated their publications through a variety of document types. There are seventeen (17) document types have identified as Article; Review; Editorial Material; Letter; Meeting Abstract; News Item; Article, Early Access; Article, Proceeding Papers; Correction; Review,

Book Chapter, Editorial Material, Early Access; Review, Early Access; Letter, Early Access; Reprint; Article, Data Paper; Editorial Material , Book Chapter

Table 6: Form Wise Distribution of Research Output

| S.No. | Document Type | Records | % | TLCS | TGCS |
|---|---|---|---|---|---|
| 1 | Article | 4648 | 76.56 | 31935 | 79257 |
| 2 | Review | 612 | 10.08 | 3479 | 12763 |
| 3 | Editorial Material | 266 | 4.38 | 553 | 1504 |
| 4 | Letter | 148 | 2.44 | 1038 | 1467 |
| 5 | Meeting Abstract | 101 | 1.66 | 1 | 6 |
| 6 | News Item | 94 | 1.55 | 65 | 126 |
| 7 | Article; Early Access | 37 | 0.61 | 0 | 53 |
| 8 | Article; Proceedings Paper | 37 | 0.61 | 112 | 339 |
| 9 | Correction | 35 | 0.58 | 49 | 68 |
| 10 | Review; Book Chapter | 33 | 0.54 | 213 | 589 |
| 11 | Article; Book Chapter | 18 | 0.30 | 83 | 284 |
| 12 | Editorial Material; Early Access | 18 | 0.30 | 0 | 3 |
| 13 | Review; Early Access | 11 | 0.18 | 0 | 3 |
| 14 | Letter; Early Access | 8 | 0.13 | 0 | 1 |
| 15 | Reprint | 3 | 0.05 | 1 | 4 |
| 16 | Article; Data Paper | 1 | 0.02 | 0 | 0 |
| 17 | Editorial Material; Book Chapter | 1 | 0.02 | 0 | 0 |
|  |  | 6071 | 100.00 | 37529 | 96467 |

**Analysis of the Publication Output of Top 20 Authors**

The ranking of authors of various research articles is displayed in Table 7 and figure 2. In the rank analysis, the authors who have published less than 43 articles were not considered into account to avoid a long list. It is observed that there are a total of 21066 authors for 6071 records and it shows the top 20 most productive authors during 2011-2020. Drosten C published 114 (1.90%) articles with 6104 TGCS articles, followed by Memish ZA 112 (1.80%) with 5445 TGCS articles, Yuen KY 104 (1.70%) with 4362 TGCS articles, Baric RS 93 (1.50%) with 3123 TGCS articles, Perlman S 85 (1.40%) with 2452 TGCS article, Woo PCY 78 (1.30%) with 2750 TGCS articles, Al-Tawfiq JA 73 (1.20%) with 3046 TGCS, Lau SKP 72 (1.20%) with 2204 TGCS and other authors have contributed less than 1.20% during the period of study. The data set clearly depicts that the number of publication by an author doesn't necessarily determine the quality of publications alone as shown in the form of total global citation score. It could be identified from author wise analysis the following authors: Drosten C, Memish ZA, Yuen KY, Baric RS, Perlman S, Woo PCY, Al-Tawfiq JA, Lau SKP, Jiang SB, and Haagmans BL are the most productive authors based on the number of research papers published in the Coronavirus research. The data set puts forth that the authors Drosten C with 6104 citations, Memish ZA with 5445 citations, Yuen KY with 4362 citations and Muller MA with 3822 citations.

Table 7: Publication output of Top 20 Authors and Citation Score

| S.No. | Authors | Records | % | TLCS | TGCS |
|---|---|---|---|---|---|
| 1 | Drosten C | 114 | 1.90 | 3581 | 6104 |
| 2 | Memish ZA | 112 | 1.80 | 3237 | 5445 |
| 3 | Yuen KY | 104 | 1.70 | 2182 | 4362 |
| 4 | Baric RS | 93 | 1.50 | 1662 | 3123 |
| 5 | Perlman S | 85 | 1.40 | 1144 | 2452 |
| 6 | Woo PCY | 78 | 1.30 | 1465 | 2750 |
| 7 | Al-Tawfiq JA | 73 | 1.20 | 1634 | 3046 |
| 8 | Lau SKP | 72 | 1.20 | 1309 | 2204 |
| 9 | Jiang SB | 68 | 1.10 | 929 | 1672 |
| 10 | Haagmans BL | 63 | 1.00 | 2002 | 3511 |
| 11 | Muller MA | 63 | 1.00 | 2419 | 3822 |
| 12 | Enjuanes L | 61 | 1.00 | 559 | 1775 |
| 13 | Du LY | 55 | 0.90 | 1003 | 1589 |
| 14 | Corman VM | 53 | 0.90 | 2424 | 3184 |
| 15 | Bosch BJ | 51 | 0.80 | 2280 | 2938 |
| 16 | Zhang Y | 51 | 0.80 | 457 | 1218 |
| 17 | Chan JFW | 45 | 0.70 | 1079 | 2196 |
| 18 | Li Y | 45 | 0.70 | 563 | 1055 |
| 19 | Chan KH | 43 | 0.70 | 1203 | 2324 |
| 20 | Gerber SI | 43 | 0.70 | 657 | 992 |

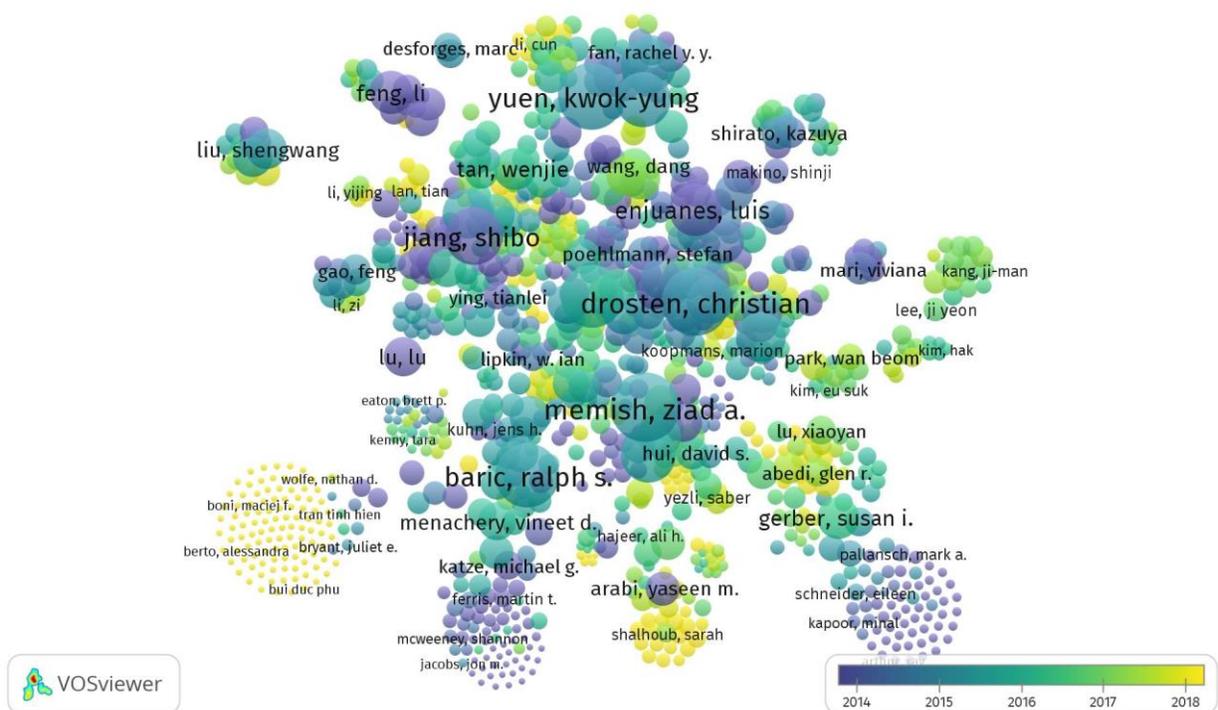

Figure 2: Highly Prolific Authors

**Analysis of the Publication Output of Top 20 Journals**

Table 8 and figure 3 displays the publication output of the top twenty journals by number of papers and Journal of Virology acquired 1st rank among the top twenty Journals under consideration with its total global citation score 9897. In all 1070 journals contributed in research during 2011 and 2020. The journals that rank between 2[nd] and 10[th] position are PLOS One, Viruses-Basel, Emerging Infectious Diseases, Virology, Virus Research, Archives of Virology, Journal of General Virology, Veterinary Microbiology, and Scientific Reports. It could be identified that the journal wise analysis the following journals: Journal of Virology, PLOS One, Virused-Basel, Emerging Infectious Veterinary Microbiology, and Scientific Reports were identified the most productive journals based on the number of research papers published.

Table 8: Distribution of the Publication Output of Top 20 Journals

| S.No. | Journals | Records | % | TLCS | TGCS |
|---|---|---|---|---|---|
| 1 | Journal of Virology | 360 | 5.90 | 5464 | 9897 |
| 2 | PLOS One | 213 | 3.50 | 0 | 3110 |
| 3 | Viruses-Basel | 168 | 2.80 | 621 | 1805 |
| 4 | Emerging Infectious Diseases | 124 | 2.00 | 3574 | 4707 |
| 5 | Virology | 124 | 2.00 | 822 | 1771 |

| 6 | Virus Research | 119 | 2.00 | 992 | 1669 |
|---|---|---|---|---|---|
| 7 | Archives of Virology | 108 | 1.80 | 501 | 977 |
| 8 | Journal of General Virology | 103 | 1.70 | 929 | 2098 |
| 9 | Veterinary Microbiology | 93 | 1.50 | 664 | 1171 |
| 10 | Scientific Reports | 82 | 1.40 | 0 | 836 |
| 11 | Virology Journal | 78 | 1.30 | 0 | 1271 |
| 12 | Antiviral Research | 73 | 1.20 | 711 | 1326 |
| 13 | Journal of Medical Virology | 69 | 1.10 | 152 | 485 |
| 14 | Journal of Virological Methods | 69 | 1.10 | 144 | 525 |
| 15 | Plos Pathogens | 63 | 1.00 | 0 | 2408 |
| 16 | MBIO | 62 | 1.00 | 0 | 2489 |
| 17 | Eurosurveillance | 60 | 1.00 | 66 | 1869 |
| 18 | Emerging Microbes & Infections | 59 | 1.00 | 34 | 523 |
| 19 | Infection Genetics and Evolution | 55 | 0.90 | 430 | 769 |
| 20 | Journal of Infectious Diseases | 54 | 0.90 | 840 | 1441 |

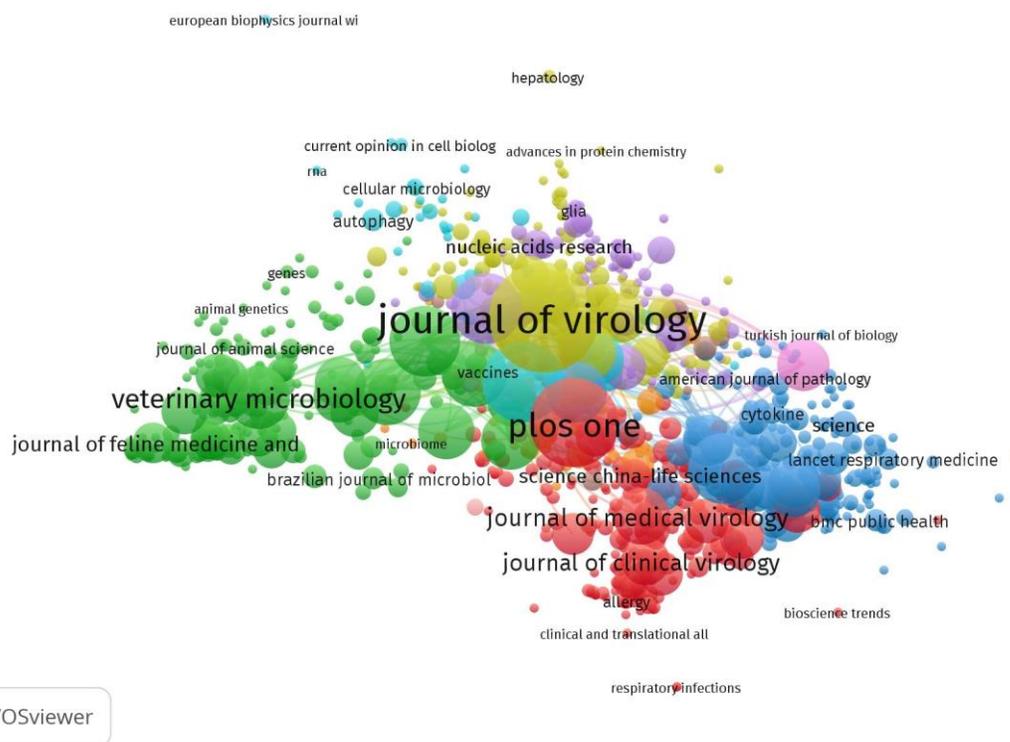

Figure 3: Publication output of Top Journals

**Analysis of the Publication Output of Top 20 Institutions**

The most prolific 20 industrious institutions were analyzed in this part. Institutions that published more than 65 and above publications have been considered as highly productive institutions.

Table 9 summarizes articles, the global citation score, local citation score and average citation per paper of the publications of these institutions. In total, 4630 institutions, including 10651 subdivisions published 6071 research papers during 2011-2020. The topmost twenty institutions involved in this research have published 65 and more research articles. The mean average is 1.31 research articles per Institution. Out of 4630 institutions, top 20 institutions published 2080 (34.26%) research papers and the rest of the institution published 3991 (65.74%) research papers respectively. Based on the number of published research records the institutions are ranked as: The institution "University of Hong Kong" holds the first rank and the institution published 236 (3.90%) research papers with 3635 local and 7436 global citation scores, the average citation per paper is 31.51. The second rank is achieved by "Chinese Academy Science " the institution published 155 (2.60%) research papers with 1718 local and 3434 global citation scores, the average citation per paper is 22.15. The "Ministry of Health" holds the 3rd rank, the institution published 149 (2.50%) research papers with 3416 local and 5503 global citation scores, and the average citation per paper is 36.93. The "Chinese Academy Agriculture Science" holds the 4th rank, the institution published 135 (2.20%) research papers with 669 local and 1714 global citation scores, the average citation per paper is 12.70. The "University Utrecht" holds the 5th rank; the institution published 109 (1.80%) research papers with 2726 local and 4298 global citation scores, the average citation per paper is 39.43. It is clear from the analysis that the following institutions: University of Hong Kong, Chinese Academy of Science, Minist Health, Chinese Academy of Agriculture Science, University of Utrecht, NIAID, Central Dis Control & Prevent, Fudan University, Erasmu MC, University of N Carolina were identified the most productive institutions based on the number of research papers published in coronavirus research. Erasmus MC (63.96), University Bonn (62.81), University Utrecht (39.43), Minst Hlth (36.93) and Leiden University (35.61) are the institutions with high ACPP score indicating the quality work with high citation impact; hence they can be recognized as the most productive institutions based on the annual citation per paper received in terms of publications.

Table 9: Ranking of Institutions and their Research Performance

| S.No. | Institution | Records | % | TLCS | TGCS | ACPP |
|---|---|---|---|---|---|---|
| 1 | University Hong Kong | 236 | 3.90 | 3635 | 7436 | 31.51 |
| 2 | Chinese Academy Science | 155 | 2.60 | 1718 | 3434 | 22.15 |
| 3 | Minist Hlth | 149 | 2.50 | 3416 | 5503 | 36.93 |
| 4 | Chinese Academy Agriculture Science | 135 | 2.20 | 669 | 1714 | 12.70 |

| 5 | University Utrecht | 109 | 1.80 | 2726 | 4298 | 39.43 |
| 6 | NIAID | 108 | 1.80 | 704 | 3258 | 30.17 |
| 7 | Ctr Dis Control & Prevent | 106 | 1.70 | 799 | 2127 | 20.07 |
| 8 | Fudan University | 106 | 1.70 | 1127 | 2074 | 19.57 |
| 9 | Erasmus MC | 100 | 1.60 | 3304 | 6396 | 63.96 |
| 10 | University N Carolina | 100 | 1.60 | 1720 | 3201 | 32.01 |
| 11 | University Bonn | 97 | 1.60 | 3453 | 6093 | 62.81 |
| 12 | University Iowa | 96 | 1.60 | 1204 | 2608 | 27.17 |
| 13 | Seoul National University | 80 | 1.30 | 546 | 1109 | 13.86 |
| 14 | University Calif Davis | 78 | 1.30 | 691 | 1678 | 21.51 |
| 15 | Al-Faisal University | 77 | 1.30 | 1033 | 1903 | 24.71 |
| 16 | University Minnesota | 73 | 1.20 | 943 | 1680 | 23.01 |
| 17 | Chinese Academy Medical Science | 72 | 1.20 | 703 | 1402 | 19.47 |
| 18 | King Saud University | 71 | 1.20 | 524 | 1213 | 17.08 |
| 19 | Leiden University | 66 | 1.10 | 357 | 2350 | 35.61 |
| 20 | University Texas Medical Branch | 66 | 1.10 | 677 | 1574 | 23.85 |

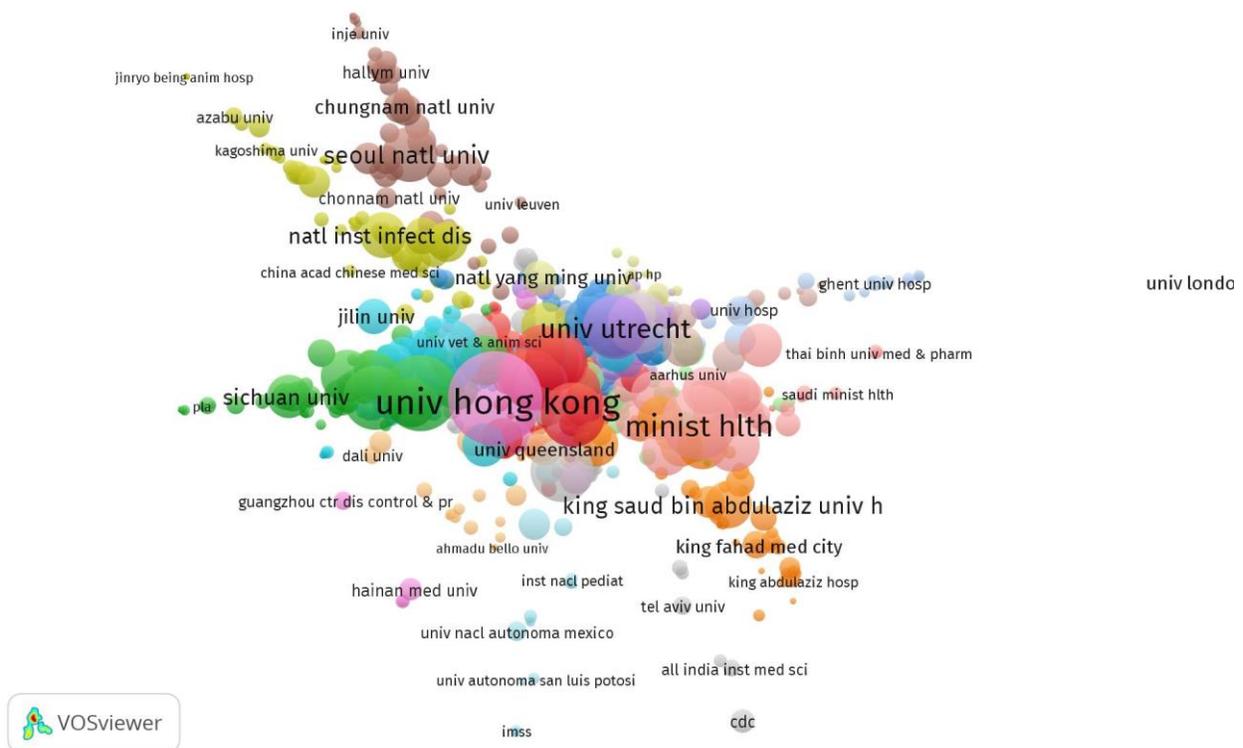

Figure 4: Collaboration of Institutions and their clusters

**Conclusion**

The number of papers published in coronavirus disease research has gradually increased during 2011–2020 and the study has shown that a total number of 6071 research documents have been published over a period of 10 years. The data from this paper also suggest that authors Drosten C, Memish ZA, Yuen KY, Baric RS, Perlman S, Woo PCY, Al-Tawfiq JA, Lau SKP, Jiang SB, and Haagmans BL, were identified as the most prolific authors based on the number of research papers contributed. It could be seen from Institutions Wise Analysis that the following institutions : University of Hong Kong, Chinese Academy of Science, Minist Health, Chinese Academy of Agriculture Science, University of Utrecht, NIAID, Central Dis Control & Prevent, Fudan University, Erasmu MC, University of N Carolina have published maximum number of research papers in the coronavirus disease research. The following countries: USA, Peoples Republic of China, UK, Saudi Arabia, Germany, South Korea, Netherlands, France, Japan, and Canada were recognised the nations that have contributed highest number of publications during the period under study. It could be identified that the journal wise analysis the following journals: Journal of Virology, PLOS One, Virused-Basel, Emerging Infectious Veterinary Microbiology, and Scientific Reports were identified the most productive journals based on the number of research papers published.

**References**

Ahmad, M., & Batcha, M. S. (2019). Mapping of Publications Productivity on Journal of Documentation 1989-2018 : A Study Based on Clarivate Analytics – Web of Science Database. *Library Philosophy and Practice (e-Journal)*, 2213–2226.

Ahmad, M., & Batcha, M. S. (2019). Scholarly Communications of Bharathiar University on Web of Science in Global Perspective: A Scientometric Assessment. *Research Journal of Library and Information Science*, *3*(3), 22–29.

Ahmad, M., Batcha, M. S., & Jahina, S. R. (2019). Testing Lotka's Law and Pattern of Author Productivity in the Scholarly Publications of Artificial Intelligence. *Library Philosophy and Practice (e-Journal)*.

Ahmad, M., Batcha, M. S., Wani, B. A., Khan, M. I., & Jahina, S. R. (2018). Research Output of Webology Journal (2013-2017): A Scientometric Analysis. *International Journal of Movement Education and Social Science*, *7*(3), 46–58.

Batcha, M. S., & Ahmad, M. (2017). Publication Trend in an Indian Journal and a Pakistan Journal: A Comparative Analysis using Scientometric Approach. *Journal of Advances in Library and Information Science*, *6*(4), 442–449.

Batcha, M. S., Dar, Y. R., & Ahmad, M. (2019). Impact and Relevance of Cognition Journal in the Field of Cognitive Science: An Evaluation. *Research Journal of Library and Information Science*, *3*(4), 21–28.

Batcha, M. S., Jahina, S. R., & Ahmad, M. (2018). Publication Trend in DESIDOC Journal of Library and Information Technology during 2013-2017: A Scientometric Approach. *International Journal of Research in Engineering, IT and Social Sciences*, *8*(04), 76–82.

Corman, V. M., Albarrak, A. M., Omrani, A. S., Albarrak, M. M., Farah, M. E., Almasri, M., … Memish, Z. A. (2015). Viral Shedding and Antibody Response in 37 Patients With Middle East Respiratory Syndrome Coronavirus Infection. *Clinical Infectious Diseases*, *62*(4), 477–483. https://doi.org/10.1093/cid/civ951

Drosten, C., Günther, S., Preiser, W., Van der Werf, S., Brodt, H. R., Becker, S., … Doerr, H. W. (2003). Identification of a novel coronavirus in patients with severe acute respiratory syndrome. *New England Journal of Medicine*, *348*(20), 1967–1976. https://doi.org/10.1056/NEJMoa030747

Grant, P. R., Garson, J. A., Tedder, R. S., Chan, P. K. S., Tam, J. S., & Sung, J. J. Y. (2003). Detection of SARS Coronavirus in Plasma by Real-Time RT-PCR. *New England Journal of Medicine*, *349*(25), 2468–2469. https://doi.org/10.1056/NEJM200312183492522

Huang, C., Wang, Y., Li, X., Ren, L., Zhao, J., Hu, Y., … Cao, B. (2020). Clinical features of patients infected with 2019 novel coronavirus in Wuhan, China. *The Lancet*, *395*(10223), 497–506. https://doi.org/10.1016/S0140-6736(20)30183-5

Ng, E. K. O., Hui, D. S., Chan, K. C. A., Hung, E. C. W., Chiu, R. W. K., Lee, N., … Lo, Y. M. D. (2003). Quantitative Analysis and Prognostic Implication of SARS Coronavirus RNA in the Plasma and Serum of Patients with Severe Acute Respiratory Syndrome. *Clinical Chemistry*, *49*(12), 1976–1980. https://doi.org/10.1373/clinchem.2003.024125

Reuters. (2020). Wuhan lockdown "unprecedented", shows commitment to contain virus: WHO representative in China. Retrieved January 23, 2020, from https://www.reuters.com/article/us-china-health-who/wuhan-lockdown-unprecedented-shows-commitment-to-contain-virus-who-representative-in-china-idUSKBN1ZM1G9

Shan, H., & Zhang, P. (2004). Viral attacks on the blood supply: The impact of severe acute respiratory syndrome in Beijing. *Transfusion*, *44*(4), 467–469. https://doi.org/10.1111/j.0041-1132.2004.04401.x

Wang, C., Horby, P. W., Hayden, F. G., & Gao, G. F. (2020). A novel coronavirus outbreak of

global health concern. *The Lancet*, *395*(10223), 470–473. https://doi.org/10.1016/S0140-6736(20)30185-9

World Health Organization. (2004). Summary of probable SARS cases with onset of illness from 1 November 2002 to 31 July 2003. Retrieved March 1, 2020, from https://www.who.int/csr/sars/country/table2004_04_21/en/

World Health Organization. (2013). Middle East respiratory syndrome coronavirus (MERSCoV). Retrieved March 1, 2020, from https://www.who.int/emergencies/mers-cov/en/

World Health Organization. (2020). Coronavirus. Retrieved March 15, 2020, from https://www.who.int/health-topics/coronavirus#tab=tab_1

World Health Organization. (2020). WHO Director-General's opening Remarks at the media briefing on COVID-19. Retrieved March 3, 2020, from https://www.who.int/dg/speeches/detail/who-director-general-s-opening-remarks-at-the-media-briefing-on-covid-19---3-march-2020

WorldoMeter. (2020). Coronavirus (COVID-19) Mortality Rate. Retrieved March 4, 2020, from https://www.worldometers.info/coronavirus/coronavirus-death-rate/